\documentstyle[11pt]{article}
\begin{document}

\def\be{\begin{equation}} \def\en{\end{equation}}

\def\B{ B}  \def\R{ R}  \def\g{\eta}  
\def\A{{\cal A}}  \def\a{{\alpha}} \def\K{{\cal H}}  \def\X{{\cal X}}
\def\F{{\cal F}} \def\P{{\cal P}} \def\Q{{\cal Q}}

\title{Amalgamated Codazzi Raychaudhuri identity for foliation.}

\author{ {\bf Brandon Carter}
\\ D.A.R.C., (UPR 176, CNRS),
\\ Observatoire de Paris, 92 Meudon, France.}

\date{ June, 1996} 

\maketitle

{\bf Abstract.\ } It is shown how a pure background tensor formalism
provides a concise but explicit and highly flexible machinery for the
generalised curvature analysis of individual embedded surfaces and
foliations such as arise in the theory of topological defects in
cosmological and other physical contexts. The unified treatment
provided here shows how the relevant extension of the Raychaudhuri
identity is related to the correspondingly extended Codazzi identity.

\section{Introduction}

I wish to thank the organisors of this Penrose festschrift meeting 
`Geometric issues in the foundations of science' for the opportunity of again
expressing appreciation for the beautifully geometric way of perceiving the 
physical world that Roger Penrose communicated to so many students of my 
generation. In particular it was Roger's emphasis\cite{[23]} on the 
importance of features that are conformally invariant that lead me, with his 
help, to develop the systematic use\cite{[24]} of 2-dimensional {\it conformal 
projections} (the Lorentz signature analogue of Mercator type projections 
in ordinary terrestrial mapping) of the kind that have since become widely 
and appropriately known as `Penrose diagrams'. An outstanding example
of the kind of conformally invariant structure whose analysis  was specially
developed and applied under Roger's leadership is that of null-geodesic
foliations: in particular, it was his derivation, with Ted Newman\cite{[25]},
of the null limit of the famous divergence identity obtained originally for a
timelike flow by Raychaudhuri\cite{[26]}, that provided the essential tool for
deriving the singularity theorems that were subsequently developed, first by
Roger himself\cite{[27]}, and later on by Stephen Hawking and
others\cite{[28],[29]}.

It was the work on singularity theorems\cite{[27],[28],[29]} that first drew
my attention to the original Raychaudhuri equation, whose extension to higher
dimensional foliations, in a manner recently suggested by Capovilla and
Guven\cite{[15]}, is described in the present article. The unified treatment
provided here shows how the extended Raychaudhuri identity is fraternally
related to the correspondingly extended Codazzi identity. However it is left
for future work to complete the corresponding Penrose program, in the sense of
treating the corresponding conformally invariant limit, meaning the case of a
foliation not by surfaces with a well behaved induced metric such as will be
postulated in the present work (which is physically motivated by contexts
such as that of neutron star vortex congruences) but by null surfaces (of the
kind whose study has been developed by Barrab\`es and Israel\cite{[30]}).

The present article is a sequel to a previous Penrose festschrift
contribution\cite{[7]} in which I showed how a pure background tensor
formalism provides a concise but explicit and highly flexible machinery for
the generalised curvature analysis of individual embedded timelike or
spacelike p-dimensional surfaces in a flat or curved n-dimensional
spacetime  background. The relevant spacetime metric will, as usual, be
denoted here by $g_{\mu\nu}$, with the understanding that the suffices are
interpretable either in the (mathematically sophisticated) sense  of Roger's
abstract index system\cite{[31]} or else in the old fashionned concrete sense
(with which most physicists are still more familiar) as the labels of
components with respect to some set of local coordinates $x^\mu$, $\mu$= 0,1,
n-1. It will be shown here how this machinery can be extended in a natural
way so as to treat a smooth foliation by a congruence of such surfaces, for
which the complete orthonormal frame bundle characterised by the relevant
group of rotations in n dimensions will have a natural reduction to the
bundle of naturally adapted frames characterised by the direct product of the
subgroup of tangential frame rotations in p dimensions and the complementary
subgroup of orthogonal frame rotations in (n-p) dimensions. This reduced
frame bundle will be naturally adowed with a preferred -- metric preserving
but generically non-symmetric -- connection, which will have an associated
{\it foliation curvature} tensor $\F_{\mu\nu\ \sigma}^{\,\ \ \rho}$ that is
generically distinct from the ordinary Riemannian {\it background curvature}
which will be denoted here by $ \B_{\mu\nu\ \sigma}^{\,\ \ \rho}$. The 
various ways in which this foliation curvature tensor can be projected, 
orthogonally or tangentially, with respect to the embedded surfaces give 
relations of which the generalised Codazzi and Raychaudhuri identities are 
particular cases.

The present approach has been developed to satisfy needs arising in the
context of the recent rise  of interest in the theory of topological  defect
structures such such as  cosmic strings and higher dimensional cosmic
membranes, as well as the related phenomenon of vortex foliations in neutron
stars, which has lead to the investigation of a wide range of new problems of
equilibrium and dynamical evolution in a special or general relativistic
framework. Various aspects of these
problems\cite{[1],[2],[3],[5],[6],[7],[8]} and in recent years most
particularly the requirements of general purpose perturbation analysis
\cite{[10],[11],[13],[14],[15],[16]}, have shown, and in some cases
helped to satisfy, the need to adapt and develop pre-existing mathematical
machinery for describing the relevant geometry and particularly the various
kinds of curvature that are involved. 

As remarked in my preceeding Penrose festchrift contribution\cite{[7]},
although much of what is needed has in principle been already available in the
mathematical literature\cite{[17],[18],[19],[20],[21]}, it has often been in a
form that is inaccessible or inconvenient for the purposes of physicists, many
of whom have remained excessively dependent on obsolete sources such as
Eisenhart's still very influential textbook\cite{[22]} (written in ignorance
of the modern concept of generalised curvature, which was at that time under
development by Cartan, and which was made familiar to physicists much later by
the theory of Yang and Mills). Whereas some treatments have obtained a neatly
concise abstract formulation at the expense of flexibility, others have
obtained general purpose adaptability at the price of using complicated and
potentially confusing reference systems involving specially adapted
coordinates and frames that require the simultaneous use of many different
kinds of indices.

For the purpose of obtaining an optimal compromise between these two
undesirable extremes, the approach\cite{[7],[8],[10],[16]} used here
relies as much as possible just on ordinary tensors, as defined with respect
to the relevant {\it background} space-time with local coordinates $x^\mu$.
The advantage of avoiding explicit dependence on a specialised internal
coordinate system becomes particularly clear in cases\cite{[8]} where one is
concerned with mutual contractions of tensors constructed on distinct but
mutually intersecting embedded surfaces whose internal coordinates could not
in general be made to be mutually compatible.

The present treatment will employ the same notation as was used in the
most mathematically detailed presentation\cite{[7]} of this approach, in 
which, rather than describing the induced curvature of an embedded spacelike
or timelike p-surface, with internal coordinates $\sigma^i$ say, in terms of 
the intrinsic version, with components $\R_{ijk\ell}$ say, of its Riemann 
tensor, one prefers to describe it in terms of the corresponding 
background spacetime tensor $\R_{\lambda\mu\nu\rho}$. The latter is 
definable (using the abbreviation $x^\mu_{\, ,i}$ for $\partial x^\mu/
\partial \sigma^i$) as the index lowered verson --  obtained by contraction
with the background spacetime metric $g_{\mu\nu}$ -- of the projection
$\R{^{\lambda\mu\nu\rho}}= \R{^{ijkl}}\,x^\lambda_{\, ,i}x^\mu_{\, ,j}
x^\nu_{\, ,k} x^\rho_{\, ,l}$ of the contravariant version of the intrinsic
curvature that is obtained by internal index raising, using the inverse, 
with contravariant components $\g^{ij}$ (whose existence depends on the 
postulate that the surface is spacelike or timelike, but not null) of the 
induced metric with components $\g_{ij}= g_{\mu\nu}x^\mu_{\, ,i} 
x^\nu_{\, ,j}$. This background spacetime representation
$\R_{\lambda\mu\nu\rho}$ of the internal curvature of the embedded
p-surface has of course to be distinguished from the ordinary n-dimensional 
Riemann curvature of the background spacetime itself, whose components 
will be denoted simply by $\B_{\lambda\mu\nu\rho}$. In the same way, rather
than working with the covariant and contravariant intrinsic 
components $\g_{ij}$ and  $\g^{ij}$ of the induced metric,  one prefers 
to use the corresponding  background coordinate components 
$\g_{\mu\nu}$ and $\g^{\mu\nu}$   of the corresponding projected tensor 
specified by $\g^{\mu\nu}= \g^{ij} x^\mu_{\, ,i}x^\nu_{\, ,j}$, which is what
is referred to as the (first) {\it fundamental tensor} of the embedding.

To set up a systematic analysis of curvature in exclusively background
tensorial terms, the natural starting point is obviously the fundamental
tensor that has just been defined, whose mixed (contra/covariant) version
with components $\g^\mu_{\ \nu}$ is interpretable as a rank-p {\it tangential
projection} operator that sends a vector into the tangent subspace of the 
embedding. The notation $\perp^{\!\mu}_{\ \nu}$ will be used here 
(instead of the less suggestive symbol $\gamma^\mu_{\ \nu}$) to denote the 
complementary rank - (n-p) operator of {\it lateral projection} orthogonal to
the surface, whose components will evidently be given in terms of those of 
the fundamental (tangential projection) tensor by the defining relation 
\be
\perp^{\!\mu}_{\ \nu}+\ \g^\mu_{\ \nu}=g^\mu_{\ \nu}\ ,
\label{1.1}
\en 
since the mixed version $g^\mu_{\ \nu}$ of the metric tensor is of course
interpretable as representing the identity operator. As well as having the
separate operator properties 
\be
\g^\mu_{\ \rho}\,\g^\rho_{\ \nu}=\g^\mu_{\ \nu} \ , \hskip 1.2 cm
\perp^{\!\mu}_{\ \rho}\perp^{\!\rho}_{\ \nu}=\perp^{\!\mu}_{\ \nu}
\label{1.2}\en 
the tensors thus defined will evidently be related by the conditions
\be
\g^\mu_{\ \rho}\perp^{\!\rho}_{\ \nu}\,=\,0\,=\perp^{\!\mu}_{\ \rho}
\g^\rho_{\ \nu} \ . 
\label{1.3}\en

\section{The deformation tensor.} 

Although it is less detailed, the present article considerably extends the
results of the preceeding development\cite{[7]} of the background tensorial
analysis of embedding geometry, by considering not just an individual embedded
non-null $p$-surface by itself, but the extension of any such surface to a
smooth foliation by diffeomorphically similar surfaces. For such a foliation
there will be corresponding {\it background} (not just single $p$-surface
supported) fields of tensors $\g^\mu_{\ \nu}$ and $\perp^{\!\mu}_{\,\nu}$,
that will not only satisfy the relations (\ref{1.1}), (\ref{1.2}), and
(\ref{1.3}), but will also (unlike what was supposed in the preceeding Penrose
festscrift article\cite{[7]}) have well defined (Riemannian or
pseudo-Riemannian) covariant derivatives. These derivatives given  will be
fully determined by the specification of a certain (first) {\it deformation
tensor}, $\K_{\mu\ \rho}^{\ \nu}$ say, via an expression of the form \be
\nabla_{\!\mu}\,\g^\nu_{\ \rho}= -\nabla_{\!\mu}\perp^{\!\nu}_{\,\rho}
=\K_{\mu\ \rho}^{\ \nu}+\K_{\mu\nu}^{\ \ \rho} \ . \label{2.1}\en It can
easily be seen from (\ref{1.2}) that the required deformation tensor will be
given simply by \be \K_{\mu\ \nu}^{\ \rho}=\g^\rho_{\
\sigma}\,\nabla_{\!\mu}\,\g^\sigma_{\ \nu}=
-\perp^{\!\sigma}_{\,\nu}\nabla_{\!\mu}\!\perp^{\!\rho}_{\,\sigma} \
.\label{2.2}\en The middle and last indices of this tensor will evidently have
the respective properties of tangentiality and orthogonality that are
expressible as \be \perp^{\!\sigma}_{\,\nu} \K_{\mu\ \sigma}^{\ \rho}= 0 \
,\hskip 1.2 cm \K_{\mu\ \sigma}^{\ \rho}\,\g^\sigma_{\ \nu}=0 \ .
\label{2.3}\en There is no automatic tangentiality or orthogonality property
for the first index of the deformation tensor (\ref{2.2}), which is thus
reducible with respect to the tangential and orthogonally lateral projections
(\ref{1.1}) to  a sum \be \K_{\mu\ \nu}^{\ \rho}=K_{\mu\ \nu}^{\
\rho}-L_{\mu\nu}^{\,\ \ \rho}\ \label{2.4}\en in which such a property is
obtained for each of the parts \be K_{\mu\ \nu}^{\ \rho}=\g^\sigma_{\ \mu}
\K_{\sigma\ \nu}^{\ \rho}\ , \hskip 1 cm L_{\mu\nu}^{\,\ \
\rho}=-\perp^{\!\sigma}_{\,\mu} \K_{\sigma\ \nu}^{\ \rho}\ , \label{2.5}\en
which satisfy the conditions \be \perp^{\!\sigma}_{\,\mu}K_{\sigma\ \nu}^{\
\rho}= \perp^{\!\rho}_{\,\sigma} K_{\mu\ \nu}^{\ \sigma}=0=K_{\mu\ \sigma}^{\
\rho}\,\g^\sigma_{\ \nu}\ . \label{2.6}\en and \be \g^{\sigma}_{\
\mu}L_{\sigma\ \nu}^{\,\ \rho}= \g^{\rho}_{\ \sigma} L_{\mu\ \nu}^{\,\
\sigma}=0=L_{\mu\ \sigma}^{\,\ \rho}\!\perp^{\!\sigma} _{\,\nu}\ .
\label{2.7}\en

It is evident  that the  first of these decomposed parts is appropriately
describable as the {\it tangential turning} tensor, since by (\ref{2.2}) and 
(\ref{2.5}) it is given by the expression
\be
K_{\mu\ \nu}^{\ \rho}=\g^\rho_{\ \sigma}\g^\tau_{\ \mu}
\nabla_{\!\tau}\g^\sigma_{\ \nu} \ , 
\label{2.8}\en
in which the only differentiation involved is contained in the tangential
gradient operator $\g^\tau_{\ \mu}\nabla_{\!\tau}$ -- which is well defined
even for fields whose support is restricted to a single embedded surface --
so that, unlike the full deformation tensor $\K_{\mu\ \nu}^{\ \rho}$ of the
foliation, the tangential turning tensor $K_{\mu\ \nu}^{\ \rho}$ is well
defined just for an individual embedded p-surface. As such, this turning
tensor  $K_{\mu\ \nu}^{\ \rho}$ is identifiable as what has been
defined\cite{[7]} as the ordinary {\it second fundamental tensor} of the
particular p-surface passing through the point under consideration. 

Up to this point, none of the relations formulated in this section depends
on the condition that the (first) fundamental tensor field $\g^\mu_{\ \nu}$
is actually tangential to well behaved p-surfaces rather than just being
an arbitrary field of rank-p projection tensors as characterised by the
purely algebraic  conditions (\ref{1.2}). As pointed out in the previous 
analysis\cite{[7]}, the Frobenius type integrability condition that is 
necessary and
sufficient for the local existence of well behaved p-surfaces tangential to
$\g^\mu_{\ \nu}$ is that the second fundamental tensor defined by (\ref{2.6})
should have the generalised Weingarten property
\be
K_{\mu\nu}^{\,\ \ \rho}=K_{(\mu\nu)}^{\,\ \ \ \rho} \hskip 1 cm
 \Leftrightarrow\hskip 1 cm
K_{[\mu\nu]}^{\,\ \ \ \rho}=0 \ ,
\label{2.9}\en
(using round and square brackets to denote index symmetrisation and
antisymmetrisation respectively) which means that it is symmetric with respect
to its two surface tangential indices. 

It is to be noticed that the second part of the decomposition (2.4), 
namely the {\it lateral turning tensor}, $L_{\mu\ \nu}^{\ \rho}$, is
expressible by the formula
\be
L_{\mu\ \nu}^{\ \rho}=\perp^{\!\rho}_{\,\sigma}\perp^{\!\tau}_{\, \mu}
\nabla_{\!\tau}\perp^{\!\sigma}_{\, \nu} \ , 
\label{2.10}\en
which differs from that in (\ref{2.8}) only by the substitution of
$\perp^{\!\nu}_{\,\mu}$  for $\g^\nu_{\ \mu}$. It evidently follows that the
necessary and sufficient integrability condition for the local existence of an
(n-p)-dimensional foliation orthogonal to the p-dimensional foliation whose
existence is guaranteed by (\ref{2.9}) is that this lateral turning tensor
should have the analogous symmetry property, which is expressible as the
vanishing of the {\it rotation tensor},  $\omega_{\mu\nu}^{\, \ \ \rho}$ say,
that is defined as its antisymmetric part in a decomposition of the form
\be
L_{\mu\nu}^{\, \ \ \rho}=\omega_{\mu\nu}^{\, \ \ \rho}+
\theta_{\mu\nu}^{\, \ \ \rho}\ ,
\hskip 1 cm \omega_{(\mu\nu)}^{\,\ \ \ \rho} =0\ ,
\hskip 1 cm \theta_{[\mu\nu]}^{\,\ \ \ \rho} =0\ .
\label{2.11}\en
The only part that remains if the foliation is (n-p) surface orthogonal is the
symmetric part, $\theta_{\mu\nu}^{\, \ \ \rho}$, which is the natural
generalisation of the two index divergence tensor $\theta_{\mu\nu}$ whose
evolution is the subject of the tensorial Raychaudhuri identity discussed by
Hawking and Ellis\cite{[29]} (the original Raychaudhuri identity\cite{[26]}
being obtained by taking the scalar trace). For a 1-dimensional timelike
foliation, which will have a unique future directed unit tangent vector
$u^\mu$, the relevant divergence tensor will be obtainable simply as
$\theta_{\mu\nu}= \theta_{\mu\nu}^{\, \ \ \rho} u_\rho$. The generalised
Raychaudhuri identity to be presented (following Capovilla and
Guven\cite{[15]}) in a later section, provides an evolution equation for the
three index generalised divergence tensor $\theta_{\mu\nu}^{\, \ \ \rho}$
which (unlike the ordinary divergence tensor $\theta_{\mu\nu}$) is always well
defined whatever the dimension of the foliation.

\section{The adapted foliation connection. }

Due to the existence of the decomposition whereby a background spacetime
vector, with components $\xi^\mu$ say, is split up by the projectors
(\ref{1.1}) as the sum of its surface tangential part $\g^\mu_{\ \nu}\,
\xi^\nu$ and its surface orthogonal part $\perp^{\!\mu}_{\,\nu} \xi^\nu$,
there will be a  corresponding adaptation of the ordinary concept of parallel
propagation with respect to the background connection $\Gamma_{\mu\ \rho}
^{\,\ \nu}$. The principle of the adapted propagation concept is to
follow up an ordinary operation of infinitesimal parallel propagation
by the projection adjustment that is needed to ensure that purely tangential
vectors propagate onto purely tangential vectors while purely orthogonal
vectors propagate onto purely orthogonal vectors. Thus for purely tangential
vectors, the effect of the adapted propagation is equivalent to that of
ordinary internal parallel propagation with respect to the induced metric
in the embedded surface, while for purely orthogonal vectors it is
interpretable as the natural generalisation of the standard concept of
Fermi-Walker propagation. For an infinitesimal displacement $dx^\mu$
the deviation between the actual component variation $(dx^\nu)\,\partial_\nu
\xi^\mu$ and the variation that would be obtained by the corresponding
adapted propagation law will be expressible in the form $(dx^\nu)\,D_\nu
\xi^\mu$ where $D$ denotes the corresponding {\it foliation adapted
differentiation} operator, whose effect will evidently be given by
\be
D_\nu\,\xi^\mu=\g^\mu_{\ \rho}\nabla_{\!\nu}\big(\g^\rho_{\ \sigma}
\xi^\sigma\big)+\perp^{\!\mu}_{\,\rho}\!\nabla_{\!\nu}\big(\!\perp^{\!\rho}
_{\,\sigma} \xi^\sigma\big) \ .
\label{4.1}\en
It can thus be seen that this operation will be expressible in the
form
\be
D_\nu\,\xi^\mu=\nabla_{\!\nu}\xi^\mu+\a_{\nu\ \sigma}^{\,\ \mu}\xi^\sigma
=\partial_\nu\,\xi^\mu+\A_{\nu\ \sigma}^{\,\ \mu}\xi^\sigma\ ,
\label{4.2}\en
where the adapted {\it foliation connection}
components $\A_{\mu\ \rho}^{\,\ \nu}$ are given by the formula
\be
\A_{\mu\ \rho}^{\,\ \nu}=\Gamma_{\mu\ \rho}^{\,\ \nu}
+\a_{\mu\ \rho}^{\,\ \nu}\ ,
\label{4.3}\en
in which the $\a_{\mu\ \rho}^{\,\ \nu}$ are the components of the relevant
{\it adaptation tensor}, whose components can be seen from 
(\ref{2.2}) to be given by
\be
\a_{\mu\nu\rho}=2\K_{\mu[\nu\rho]} \ .
\label{4.5}\en

The fact that the expression (\ref{4.5}) is manifestly antisymmetric with
respect to the last two indices of the adaptation tensor makes it
evident that, like the usual Riemannian differentiation operator $\nabla$,
the adapted differentiation operator $D$ will commute with index
raising or lowering, since the metric itself remains invariant under adapted
propagation:
\be
D_\mu\, g_{\nu\rho}=0\ .
\label{4.7}\en
However, unlike $\nabla$, the adapted differentiation operator has the
very convenient property of also commuting with tangential and orthogonal
projection, since it can be seen to follow from (\ref{2.1}) 
and (\ref{2.3}) that
the corresponding operators also remain invariant under adapted propagation:
\be
D_\mu\, \g^\nu_{\ \rho}=0\ ,\hskip 1 cm D_\mu\!\perp^{\!\nu}_{\,\rho}=0
\ .
\label{4.8}\en

There is of course a price to be paid in order to obtain this considerable
advantage of $D$ over $\nabla$, but it is not exhorbitant: all that has to 
be sacrificed is the analogue of the symmetry property 
\be
\Gamma_{[\mu\ \rho]}^{\ \ \nu}=0 \ .
\label{4.9}\en 
expressing the absence of torsion in the Riemannian case. For the adapted
foliation connection $\A_{\mu\ \rho}^{\,\ \nu}$, the torsion tensor defined by
\be
\Theta_{\mu\ \rho}^{\,\ \nu}=2\A_{[\mu\ \rho]}^{\ \ \nu}=
2\a_{[\mu\ \rho]}^{\ \ \nu}\ ,
\label{4.10}\en
will not in general be zero.

\section{ The amalgamated foliation curvature tensor.}

The curvature associated with the adapted connection introduced by (4.2) in
the preceeding section can be read out from the ensuing commutator formula,
which, for an abitrary vector field with components $\xi^\mu$, will take the
standard form
\be
2D_{[\mu}D_{\nu]}\xi^\rho=\F_{\mu\nu\ \sigma}^{\,\ \ \rho}\xi^\sigma
-\Theta_{\mu\ \nu}^{\,\ \sigma}D_\sigma\xi^\rho\ ,
\label{5.1}\en
in which the torsion tensor components $\Theta_{\mu\ \nu}^{\,\ \sigma}$ are 
as defined by (\ref{4.10}) 
while the components $\F_{\mu\nu\ \sigma}^{\,\ \ \rho}$ 
are defined by a Yang-Mills type curvature formula of the form
\be
\F_{\mu\nu\ \sigma}^{\,\ \ \rho} =2\partial{_{[\mu}}\A{_{\nu]}}
^\rho{_\sigma}+2\A{_{[\mu}}^{\rho\tau}\A_{\nu]\tau\sigma} \ .
\label{5.2}\en
Although the connection components $\A_{\mu\ \rho}^{\,\ \nu}$ from which it
is constructed are not of tensorial type, the resulting curvature components
(\ref{5.2}) are of course strictly tensorial. 
This is made evident by evaluating
the components (\ref{5.2}) of this {\it amalgamated foliation curvature} 
in terms of the background curvature tensor 
\be
\B_{\mu\nu\ \sigma}^{\,\ \ \rho}=2\partial {_{[\mu}}\Gamma{_{\nu]}}
^\rho{_\sigma}+\Gamma_{\mu\ \tau}^{\ \rho}\Gamma_{\nu\ \sigma}^{\ \tau}- 
\Gamma_{\nu\ \tau}^{\ \rho}\Gamma_{\mu\ \sigma}^{\ \tau}\ ,
\label{5.3}\en 
and the adaptation tensor $\a_{\mu\ \rho}^{\,\ \nu}$ given by (\ref{4.5}), 
which gives the manifestly tensorial expression
\be
\F_{\mu\nu\ \sigma}^{\,\ \ \rho} = \B_{\mu\nu\ \sigma}^{\,\ \ \rho}+
2\nabla{_{\![\mu}}\a{_{\nu]}}^\rho{_\sigma} 
+2\a{_{[\mu}}^{\rho\tau}\a{_{\nu]\tau\sigma}} \ .
\label{5.4}\en

Although it does not share the full set of symmetries of the Riemann 
tensor, the foliation curvature obtained in this way will evidently be 
antisymmetric in both its first and last pairs of indices:
\be
\F_{\mu\nu\rho\sigma}=\F_{[\mu\nu][\rho\sigma]} \ .
\label{5.5}\en
Using the formula (\ref{4.5}), it can be see from (\ref{2.3}) that the
difference between this adapted curvature and the ordinary background Riemann
curvature will be given by
\be
\F_{\mu\nu}^{\,\ \ \rho\sigma} - \B_{\mu\nu}^{\,\ \ \rho\sigma}
=4\X_{[\mu\nu]}{^{[\rho\sigma]}} +2\K_{[\mu}^{\,\ \sigma\tau}
\K_{\nu]\ \tau}^{\,\ \rho} +2\K_{[\mu}^{\,\ \tau\rho}
\K_{\nu]\tau}^{\,\ \ \sigma}  \ ,
\label{5.6}\en
where $\X_{\lambda\mu\ \rho}^{\ \ \nu}$ is what may be termed the {\it second
deformation tensor}, which is definable by
\be
\X_{\lambda\mu\ \rho}^{\ \ \nu}=
\g^\nu_{\ \sigma}\perp^{\!\tau}_{\,\rho}\nabla_{\!\lambda}\,\K_{\mu\ \tau}
^{\ \sigma} \ .
\label{3.2}\en
The formula (\ref{5.6}) superficially appears to depend on the higher order
derivatives involved in $\X_{\mu\nu}^{\ \ \rho\sigma}$, but this is deceptive:
the higher derivatives will in fact cancel, by the ``amalgamated
Codazzi-Raychaudhuri identity'' given below. 

Since the adapted derivation operator has been constructed in such a way as to
map tangential vector fields into purely tangential vector fields, and lateral
(surface orthogonal) vector fields into lateral vector fields, it follows that
the same applies to the corresponding curvature (\ref{5.2}), which will
therefore consist of an additive amalgamation of two separate parts having the
form
\be
 \F_{\mu\nu\ \sigma}^{\,\ \ \rho}=\P_{\mu\nu\  \sigma}^{\,\ \ \rho}
+\Q_{\mu\nu\ \sigma}^{\,\ \ \rho}\ ,
\label{5.7}\en
in which the ``inner" curvature acting on purely tangential vectors 
is given by a doubly tangential (surface parallel) projection as
\be
\P_{\mu\nu\ \sigma}^{\,\ \ \rho}=\F_{\mu\nu\ \lambda}^{\,\ \ \kappa}\,
\g^\rho_{\ \kappa}\,\g^\lambda_{\ \sigma} \ , 
\label{5.8}\en
while  the ``outer" curvature acting on purely orthogonal vectors is given 
by a doubly lateral projection as
\be
\Q_{\mu\nu\ \sigma}^{\,\ \ \rho}=\F_{\mu\nu\ \lambda}^{\,\ \ \kappa}
\!\perp^{\!\rho}_{\,\kappa} \perp^{\!\lambda}_{\,\sigma}\ .
\label{5.9}\en
It is implicit in the separation expressed by (\ref{5.7}) that the mixed
tangential and lateral projection of the adapted curvature must vanish:
\be
\F_{\mu\nu\ \lambda}^{\,\ \ \kappa}\,\g^\rho_{\ \kappa}
\!\perp^{\!\lambda}_{\, \sigma}=0\ .
\label{5.10}\en 

To get back, from the extended foliation curvature tensors that have just
been introduced, to their antecedent analogues\cite{[7]} for an individual
embedded surface, the first step is to construct the {\it amalgamated
embedding curvature} tensor, 
$F_{\mu\nu\ \sigma}^{\,\ \ \rho}$ say,
which will be obtainable from the corresponding amalgamated foliation
curvature $\F_{\mu\nu\ \sigma}^{\,\ \ \rho}$ by a doubly tangential
projection having the form
\be
F_{\mu\nu\ \sigma}^{\,\ \ \rho}=\g^\alpha_{\ \mu}\,\g^\beta_{\ \nu}                                
\F_{\alpha\beta\ \sigma}^{\,\ \ \rho}\ .
\label{6.1}\en
As did the extended foliation curvature, so also  this amalgated embedding
curvature will separate as the sum of ``inner" and ``outer" parts in the 
form
\be
F_{\mu\nu\ \sigma}^{\,\ \ \rho}=\R_{\mu\nu\  \sigma}^{\,\ \ \rho}
+\Omega_{\mu\nu\ \sigma}^{\,\ \ \rho}\ ,
\label{6.2}\en
in which the ``inner" embedding curvature is given by another doubly 
tangential projection as
\be
\R_{\mu\nu\ \sigma}^{\,\ \ \rho}=F_{\mu\nu\ \lambda}^{\,\ \ \kappa}\,
\g^\rho_{\ \kappa}\,\g^\lambda_{\ \sigma} 
=\g^\alpha_{\ \mu}\g^\beta_{\ \nu}\P_{\alpha\beta\ \sigma}^{\,\ \ \rho}
\ , 
\label{6.3}\en
while  the ``outer" embedding curvature (whose noteworthy property
of {\it conformal invariance} was pointed out in my previous Penrose 
festschrift contribution\cite{[7]}) is given by the corresponding 
doubly lateral projection as
\be
\Omega_{\mu\nu\ \sigma}^{\,\ \ \rho}=F_{\mu\nu\ \lambda}^{\,\ \ \kappa}
\!\perp^{\!\rho}_{\,\kappa} \perp^{\!\lambda}_{\,\sigma}
=\g^\alpha_{\ \mu}\g^\beta_{\ \nu}\Q_{\alpha\beta\ \sigma}^{\,\ \ \rho}
\ .
\label{6.4}\en

The formula (\ref{5.6}) can  be used to evaluate the ``inner" tangential part
of the foliation curvature tensor as 

\be
\P_{\mu\nu\ \sigma}^{\,\ \ \rho} = 2\K_{[\nu}{^{\rho\tau}}
\K_{\mu]\sigma\tau} +\B_{\mu\nu\ \lambda}^{\,\ \ \kappa}\,
 \g^\rho_{\ \kappa}\,\g^{\lambda}_{\ \sigma}
\label{5.12}\en
and to evaluate the ``outer" orthogonal part of the foliation curvature
tensor as
\be
\Q_{\mu\nu\ \sigma}^{\,\ \ \rho} = 2\K_{[\mu}{^{\tau\rho}}
\K_{\nu]\tau\sigma} + \B_{\mu\nu\ \lambda}^{\,\ \ \kappa}\!
 \perp^{\!\rho}_{\,\kappa}\!\perp^{\!\lambda}_{\,\sigma} \ .
\label{5.13}\en

The formula (\ref{5.12}) for the ``inner" foliation curvature is evidently
classifiable as an further extension of the previously derived
generalisation\cite{[7]} for $\R_{\mu\nu\ \sigma}^{\,\ \ \rho}$ of the
historic Gauss equation. Similarly the formula (\ref{5.13}) for the ``outer"
foliation curvature is an analogous extension of the relation\cite{[7]} for
$\Omega_{\mu\nu\ \sigma}^{\,\ \ \rho}$ that corresponds to what has sometimes
been referred to as the ``Ricci equation" but what would seem more
appropriately describable  as the {\it Schouten equation}, with reference to
the earliest relevant source with which I am familiar\cite{[17]}, since long
after the time of Ricci it was not yet understood even by such a leading
geometer as Eisenhart\cite{[22]}.

In much the same way, the non-trivial separation identity (\ref{5.10}) can be
considered as a generalisation to the case of foliations of the relation that
is itself interpretable as an extended generalisation to higher dimensions of
the historic Codazzi equation that was originally formulated in the restricted
context of 3-dimensional flat space. It can be seen from (\ref{5.6}) that this
extended Codazzi identity is expressible as 
\be 
2\X_{[\mu\nu]}{^\rho_{\ \sigma} }+\B_{\mu\nu\ \lambda}^{\,\ \ \kappa}\,
\g^\rho_{\ \kappa}\!\perp^{\!\lambda}_{\,\sigma} = 0\ , 
\label{5.11}\en 
which shows that the relevant higher derivatives are all determined entirely
by the Riemannian background curvature so that no specific knowledge of the
second deformation tensor is needed. By doubly tangential projection of the
first two indices, the extended generalisation (\ref{5.11}) will give back the
already familiar version\cite{[7]} of the generalised Codazzi identity for the
individual embedded p-surfaces of the foliation. The corresponding doubly
lateral projection would give the analogous result for the orthogonal
foliation by (n-p)-surfaces that would exist in the irrotational case for
which the lateral turning tensor given by (\ref{2.10}) is symmetric. Finally
the corresponding mixed tangential and lateral projection of (\ref{5.11})
gives an identity that is expressible in terms of foliation adapted
differentiation (\ref{4.2}) as
\be
\perp^{\!\tau}_{\,\mu}\!D^{\,}_{\!\tau} K_{\nu\ \sigma}^{\ \rho}
+\g^\tau_{\ \nu}D^{\,}_{\!\tau} L_{\mu\sigma}^{\,\ \ \rho}
=K_{\nu\ \mu}^{\ \lambda}K_{\lambda\ \sigma}^{\ \rho}+L_{\mu\ \nu}^{\ \lambda}
L_{\lambda\sigma}^{\ \ \rho}+\g^\alpha_{\ \nu}\!\perp^{\!\beta}_{\,\mu}\!
\B_{\alpha\beta\ \lambda}^{\,\ \ \kappa}\,
\g^\rho_{\ \kappa}\!\perp^{\!\lambda}_{\,\sigma}\ .
\en
This last result is interpretable as the translation into the pure background
tensorial formalism used here of the recently derived
generalisation\cite{[15]} to higher dimensional foliations of the well known
Raychaudhuri equation (whose original scalar version\cite{[26]}, and its
tensorial extension\cite{[29]}, were formulated just for the special case of a
foliation by 1-dimensional curves). The complete identity (\ref{5.11}) is
therefore interpretable  as an amalgamated  Raychaudhuri-Codazzi identity.

 \end{document}